\documentclass[10pt]{article}
\usepackage[margin=1in]{geometry} 
\usepackage{tikz}
\usepackage{authblk}

\usetikzlibrary{decorations.pathmorphing} % For springs
\usetikzlibrary{positioning} % For easy node placement
\usepackage{circuitikz}
\usepackage{enumitem}
\usepackage{amsmath}
\usepackage{amsfonts}
\usepackage{graphicx}
\usepackage{subcaption}
\usepackage{float} 
\usepackage{algorithm}
\usepackage{algorithmic}
\usepackage{cuted}
\usepackage{multicol}
\usepackage[british]{babel}  % or \usepackage[UKenglish]{babel}
\usepackage[numbers,sort&compress]{natbib}
\usepackage{titlesec}
\titleformat{\section}[block]{\normalfont\bfseries}{}{0pt}{}  % Remove section number
\titleformat{\section}
  {\normalfont\Large\bfseries}
  {} % No numbering
  {0pt}
  {}
\titlespacing*{\section}
  {0pt}{5pt}{5pt}

\titleformat{\subsection}
  {\normalfont\bfseries}
  {} % No numbering
  {0pt}
  {}
\titlespacing*{\subsection}
  {15pt}{5pt}{0pt}

\title{Van der Waals-Driven Network Restructuring Explains Time-Dependent Piezoresistivity in Soft Nanocomposites}

\author[1]{Logan Ritchie}
\author[2,3]{Elke Pahl}
\author[1]{Iain Anderson}

\affil[1]{Biomimetics Laboratory, Auckland Bioengineering Institute, The University of Auckland, 
Auckland, New Zealand}
\affil[2]{MacDiarmid Institute for Advanced Materials and Nanotechnology, Wellington, New Zealand}
\affil[3]{Department of Physics, The University of Auckland, Auckland, New Zealand}

\date{}  % Leave date blank

\begin{document}

% Generate title page
\maketitle
\begin{abstract}
Carbon-elastomer composites exhibit complex piezoresistive behaviour that cannot be fully explained by existing macroscopic or microstructural models. In this work, we introduce a network-based modelling methodology to explore the hypothesis that van der Waals interactions between carbon particles contribute to the formation of a conductivity-promoting network structure prior to curing. We combine a discrete aggregate-based representation of filler with a mesh-free, quasi-static viscoelastic model adapted from bond-based peridynamics, resolving equilibrium states through energy minimization. The resulting particle networks are analysed using graph-theoretic measures of connectivity and conductivity. Our simulations reproduce several unexplained experimental phenomena, including long-timescale resistivity decay, non-monotonic secondary peaks upon strain release, and the increasing prominence of these features with higher filler density. Crucially, these behaviours emerge from the interplay between viscoelastic stresses and van der Waals interactions. We show that the resistance response of the network operates over different characteristic timescales to the viscoelastic stress response. The approach has potential for understanding and predicting emergent behaviour in composite materials more broadly, where material characteristics often depend on percolating network structure.
\end{abstract}

% Add your content here
\begin{multicols}{2}
\section{Introduction}\label{sec:Introduction}
By adding to an insulating elastomer a sufficient quantity of conductive carbon filler, a continuously connected network of particles (percolating nanoscale network or PNN) forms within the elastomer \cite{BALBERG2002acomprehensive}, conducting electricity through tunnelling conduction between particles \cite{zhang2007carbon,Balberg2020thephysical}. Deformation of such composites changes the distribution of interparticle distances, altering the overall electrical resistivity \cite{Aviles2018piezoresistivity}. This piezoresistivity is important for a range of applications in sensing, conduction and switching for wearables and soft robotics \cite{DUAN2020recent,Wang2019advanced,henke2018modeling}.

\begin{figure}[H]
    \centering
    \includegraphics[width=0.9\linewidth]{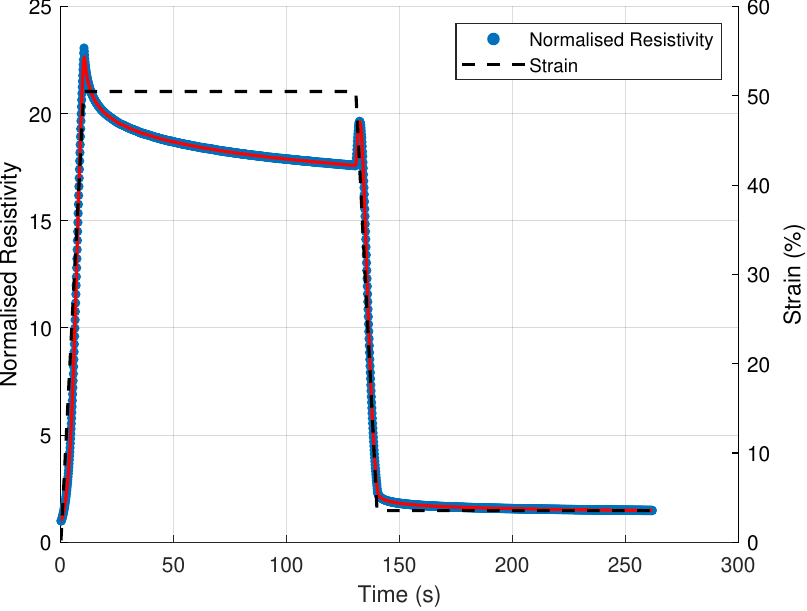}
    \caption{Piezoresistive behaviour of a silicone/carbon black composite, tested as described in \cite{ritchie2024electromechanical}, displaying time dependence and non-monotonic secondary peak of resistivity}
    \label{fig:example_piezoresistive_response}
\end{figure}

Common carbon fillers for these composites are carbon black, carbon nanotubes, and graphene. All exhibit several unexplained electromechanical behaviours \cite{Mersch2023properties,ritchie2024electromechanical,yong2022modeling,wiessner2020piezoresistivity,Tairych2019capacitive}, including significant resistance decay over timescales longer than the viscoelastic stress response, non-monotonicity and transient peaks in resistivity, and growing significance of non-monotonicity with increasing filler density. An example of this behaviour is shown in Figure~\ref{fig:example_piezoresistive_response}.

While plausible as a mechanism, a rigorous description of how changes to interparticle distances relate to conductivity is challenging to produce, as the network structure is stochastic and aperiodic. Many successful theoretical approaches to composite materials rely on assumptions of homogeneity on large scales, despite heterogeneous microstructure \cite{Choy2015Essentials}. Percolation theory \cite{stauffer2018introduction}, and effective media theories \cite{mclachlan1990electrical,Choy2015Essentials}, for example, can describe the relationship between conductivity and filler density. However, as the piezoresistive effect depends on changes in the filler distribution, it is not amenable to approaches that rely on an assumption of homogeneous distribution.

Many researchers have proposed new models to better describe this effect, broadly placed into two categories:
\begin{enumerate}[label=\roman*.,nosep] 
    \item Macroscopic, phenomenological models.
    \item Microstructural models that consider the individual filler particles.
\end{enumerate}

Several phenomenological models have reproduced the complexity of this effect by describing the relationship between resistance and strain with viscoelastic models normally used to model the stress-strain relationship \cite{Mersch2023properties,wiessner2020piezoresistivity,yong2022modeling}. These highlight the relationship between the viscoelastic properties of the elastomer and the changes in resistance during deformation, but they do not explain the underlying physical mechanism.

Computational, microstructural models often require significant simplifying assumptions in order to render the problem tractable. A variety of models are available in the literature, including finite element models \cite{Haghgoo2024breadth,WANG2018theelectro}, as well as custom computational models that investigate the role of filler reorientation under strain \cite{GONG2014carbon,Haghgoo2023analytical,Haghgoo2020acomprehensive,hu2012multi,shin2023theoretical,gbaguidi2018monte}. But limitations of these models prevent them from being able to elucidate a generic mechanism through which the full range of experimental features can be reproduced, including:
\begin{itemize}[nosep]
    \item Unrealistic Poisson ratios less than 0.5.
    \item Neglect of filler interactions.
    \item Neglect of viscoelastic matrix properties.
\end{itemize}

Furthermore, much experimental, macro-scale characterisation work \cite{Aviles2018piezoresistivity,Mersch2023properties} frequently refers to the broad concept of rearrangement of the percolating network as the cause of piezoresistivity. By contrast, much of the microstructural modelling work is focused on the use of detailed tunnelling conductivity models such as the Simmons approximation \cite{Simmons1963generalized}, and less consideration of the interparticle interactions and network restructuring.

Graph theoretic approaches have emerged as a powerful tool to describe the structural complexity of this class of materials, which exhibit `non-random disorder' \cite{Mingqiang2020biomorphic,Simoes2010applications,Vecchio2021structural,mao2024complexity}. By representing the percolating network as a graph of nodes connected by edges, network properties such as clustering and connectivity can be determined, and correlated with macro-scale material properties. 

The piezoresistive effect in conductive composites represents an ideal case study for the application of these new methodologies. It is easily measured and observed in experiment, yet the underlying complexity is considerable, as it depends not merely on a single, static network configuration, but also on the evolution of the network under deformation and over time.

This work will introduce a custom network-based modelling methodology to examine the plausibility of a simple hypothesis: Van der Waals attractions between carbon particles lead to the creation of a conductivity-promoting structure prior to curing. Deformation of the composite disrupts this favourable structure, decreasing conductivity. As the viscoelastic matrix surrounding the filler relaxes, van der Waals forces produce a partial reformation of this structure over time, partially restoring conductivity. The modelling in this work is focused on carbon black, but may be extended to include other carbon fillers such as nanotubes or graphene.
 
\section{Methods}\label{sec:modelling}
This modelling approach combined a discrete representation of the aggregates with a mesh-free, non-local formulation of continuum mechanics adapted from bond-based peridynamics \cite{SILLING2000reformulation}. Energy minimization was used to resolve the quasi-static equilibrium state of the filler network, and two graph theoretic measures were used to describe the connectivity and conductivity of the network. Due to computational limitations, the modelling presented here has been carried out in 2D.

The `diffusion limited aggregation' algorithm was used to generate rigid aggregates of multiple particles \cite{witten1983diffusion}, which represented the discrete units of the model. 

A random sequential adsorption algorithm was used to generate a random distribution of aggregates \cite{WANG2018theelectro}. This algorithm sequentially generated aggregates with a random position and orientation, and rejected the aggregate if it overlapped with any existing aggregates, resulting in a random but non-overlapping distribution.

To model pairwise particle interactions, an undirected graph was created, with individual particles as the nodes, and interactions between particles as the edges. Edges were not formed between particles that belong to the same rigid aggregate, as they are rigidly bonded.

To describe the interaction between two particles, the Everaers potential was used \cite{Everaers2003interaction}. This describes both the attractive van der Waals interaction between separated particles, and the rigid repulsion when particles come close to overlapping (Equation~\ref{eqn:veraers Potential}).
\begin{equation}
\begin{split}
U_\text{vdw} &= \frac{A}{37800}\frac{\sigma_{LJ}^6}{r} 
    \left( 
    \frac{r^2 - 14Rr + 54R^2}{(r-2R)^7} 
    \right. \\  % Invisible delimiter here
    &\quad \left. + \frac{r^2 + 14Rr + 54R^2}{(r+2R)^7} 
    - \frac{2r^2 - 60R^2}{r^7} \right) \\
&\quad - \frac{A}{6} 
    \left( 
    \frac{2R^2}{r^2 - 4R^2} 
    + \frac{2R^2}{r^2} + \ln\frac{r^2-4R^2}{r^2}
    \right)
    \label{eqn:veraers Potential}
\end{split}
\end{equation}

Where $U_{\text{vdw}}$ is the van der Waals potential energy, \textit{A} is the Hamaker constant, $\sigma_{LJ}$ is the characteristic Lennard Jones length, \textit{r} is the distance between particle centres, and \textit{R} is the radius of the particles.
By minimising the total interaction energy of the network with respect to aggregate positions and orientation, the equilibrium structure formed due to particle interactions was found. In this work, the nearly-exact trust region method was used to perform the minimization, and the minimization proceeded until the Euclidean norm of the gradient vector fell below a defined tolerance \textit{gtol} \cite{nocedal1999numerical}.
To examine the types of structures formed due to van der Waals interactions alone, the following system was simulated:
\begin{figure}[H]
\centering
\noindent\textbf{System 1}
\begin{itemize}[nosep]
    \item Particle Radius $R=50$~nm
    \item 25 Aggregates of 15 particles
    \item 25 Aggregates of 5 particles
    \item Simulation region 3900~nm\textsuperscript{2}
    \item Hamaker constant $A=1\times10^{-19}$~J
    \item Maximum interaction distance 800~nm
    \item Periodic boundary conditions
\end{itemize}
\end{figure}

The electrical network of filler particles was also described as an undirected weighted graph. Each aggregate was represented as a single node, and inter-aggregate connections are the edges, where the edge weights represent the conductivity between two aggregates. 
Multiple connections between an aggregate pair were combined into a single, effective conductance according to:
\begin{equation}
    g_{ij}=\sum_kg_k
\end{equation}
Where $g_{ij}$ is the effective conductance between aggregate $i$ and $j$, and $g_k$ is the conductance of each individual connection between them. For most simulations $g_k$ was set to a constant of 1 for all connections, and connections were created between any two particles with surface separations less than 3~nm, a reasonable maximum distance for electron tunnelling \cite{Aviles2018piezoresistivity}.

Two measures were used to describe the connectivity and conductivity of the entire network. For describing the connectivity of the network, the weighted average nodal degree was used:
\begin{equation}    
    \overline{k} =\frac{\displaystyle\sum_{i,<j>}g_{ij}}{n}
\end{equation}

Where $\overline{k}$ is the weighted average nodal degree, which served as a simple measure of the connectivity of the graph. 

To describe the spatial conductivity/resistivity of the network, the following measure was used:
\begin{subequations}
    \begin{equation}
        \overline{\sigma}=\sum_{i<j}\sigma_{ij}r_{ij}
    \end{equation}
    \begin{equation}
        \overline{\rho}=\frac{1}{\overline{\sigma}}
    \end{equation}
\end{subequations}

Where $\overline{\sigma}$ and $\overline{\rho}$ represent the measures of conductivity and resistivity, respectively, $\sigma_{ij}$ is the effective conductance between nodes \textit{i} and \textit{j}, and $r_{ij}$ is the absolute distance between nodes \textit{i} and \textit{j}. This measure represents the conductance between every pair of nodes ($i,j$) in the network, weighted by distance. 

$\sigma_{ij}$ was calculated from the effective resistance between two nodes as described in \cite{Ghosh2008minimizing}.

To simulate the effect of the viscoelastic elastomer matrix on the filler network, the elastomer was represented by one-dimensional linear viscoelastic bonds between the filler particles. This approach is an adaptation of bond-based peridynamics \cite{SILLING2000reformulation}. 

The viscoelastic bonds were represented by the Prony series \cite{Abaqus2024theory}, which represents the relaxation of the elastic modulus of a material through a series of exponentials:
\begin{equation}
    G(t)=G_\infty+\sum_iG_i\text{exp}\left(\frac{-t}{\tau_i}\right)
\end{equation}
Where \textit{G} is the current modulus, $G_\infty$ is the long term modulus, $G_i$ is the modulus of term \textit{i}, and $\tau_i$ is the relaxation time of term \textit{i}. Also useful to define is the instantaneous modulus $G_0=G_\infty+\sum_iG_i$ and the relative modulus of each term $\alpha_i=G_i/G_0$.
Following the procedure in \cite{Abaqus2024theory}, with discrete timesteps, the instantaneous elastic stresses in each bond can be calculated.
The current elastic strain energy $U_{\text{elastic}}$ for a linear viscoelastic element is defined as \cite{Abaqus2024theory}:
\begin{subequations}
    \begin{equation}
        W_{\text{elastic}}=\frac{1}{2}\sigma\varepsilon
    \end{equation}
    \begin{equation}
        U_{\text{elastic}} = W_{\text{elastic}}r_0
    \end{equation}
\end{subequations}
Where $W_{\text{elastic}}$ is the elastic strain energy density, $\varepsilon$ is strain, $\sigma$ is stress, and $r_0$ is the equilibrium length of the bond.

Thus, the total potential energy at each timestep was the sum of the elastic strain energy of each bond, and the interparticle interactions described previously.
\begin{equation}
    U_\text{total} = U_\text{vdw} + U_\text{elastic}
\end{equation}
By minimizing $U_\text{total}$, the quasi-static equilibrium structure was resolved as the network was deformed, accounting for both interparticle interactions and the viscoelastic matrix.

To investigate the time-dependent piezoresistive behaviour of the filler network, strain simulations were carried out after determining the initial network structure formed under van der Waals interactions. Post-equilibration, the viscoelastic bonds were created and initialised with equilibrium lengths equal to the current distance between bonded particles. The same parameters as System 1 were used, but the simulation region was varied between 3200~nm\textsuperscript{2} and 4150~nm\textsuperscript{2}, to test four different filler densities by area (22\%, 26\%, 30\%, 38\%). The straining procedure was as follows:
\vspace{0.5em}

{
\centering
\noindent\textbf{Strain simulation}\label{strain_sim}
\begin{itemize}[nosep]
    \item Total uniaxial strain (isochoric): 50\%.
    \item Timestep size: 1~s.
    \item Increasing strain timesteps: 20.
    \item Static holding timesteps: 120.
    \item Decreasing strain timesteps: 20.
    \item Static holding timesteps: 120.
    \item Viscoelastic elements: $G_\infty=1$~MPa, $G_1=10$~MPa, $\tau_1=1$~s
\end{itemize}
\vspace{0.5em}
}

At each timestep, the potential energy was minimized to determine the quasi-static equilibrium network structure, and the electrical properties calculated. Each simulation was averaged over 20 runs with different random seeds to reduce noise. Energy minimization was performed using the nearly exact trust region algorithm available in SciPy \cite{2020SciPy-NMeth}. This required derivation of the analytical Jacobian and Hessian, provided in the Supplementary Material.

To compare the time dependent properties of the stress response and piezoresistive response, the total viscoelastic stress was calculated by summing the absolute stresses over all bonds. The modelling software Abaqus\textsuperscript{TM} was then used to fit the Prony series to both the stress-time and resistivity-time data, allowing for a comparison between the magnitude ($\alpha_i$) and characteristic timescales ($\tau_i$) of the time dependent component of the stress and resistivity responses.

To investigate the significance of the inter-particle tunnelling conduction model, four different thresholds (0.6~nm, 3~nm, 10~nm, 100~nm) were used to define conductance between particles. Also tested was replacing the fixed bond conductance $g_k$ with the distance-dependent Simmons tunnelling approximation (Equation~\ref{eqn:simmons}) \cite{Simmons1963generalized}.
\begin{equation}
    g(d)=\frac{e^2\sqrt{2m\Delta E}}{h^2}\exp\left(-\frac{4\pi d}{h}\right)
    \label{eqn:simmons}
\end{equation}
Where $\rho$ is interparticle resistivity, $h$ is Planck's constant, $e$ and $m$ are the electron charge and mass, respectively, $\Delta E$ is the tunnelling energy barrier, and $d$ is the separation between surfaces.

\section{Results}\label{sec:Results}
\begin{figure}[H]
    \centering
    \begin{subfigure}[b]{0.4\textwidth}
        \centering
        \includegraphics[width=\textwidth]{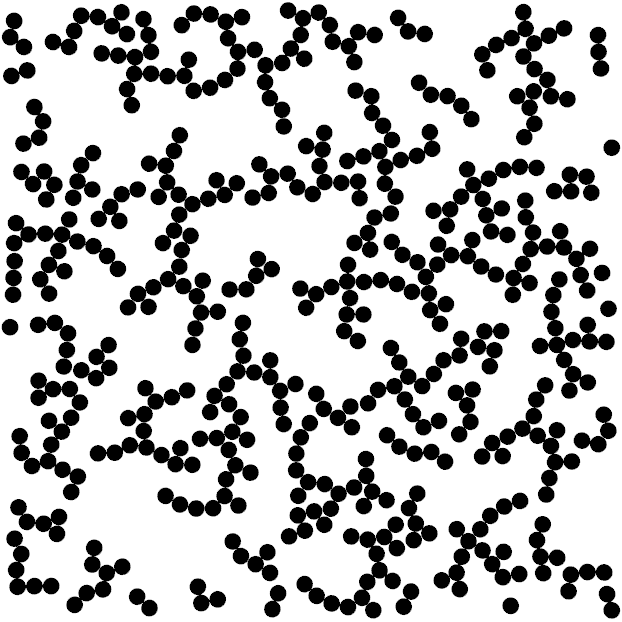}
        \caption{Initial state}
        \label{subfig:min_initial}
    \end{subfigure}
    \hspace{0.05\textwidth}
    \begin{subfigure}[b]{0.4\textwidth}
        \centering
        \includegraphics[width=\textwidth]{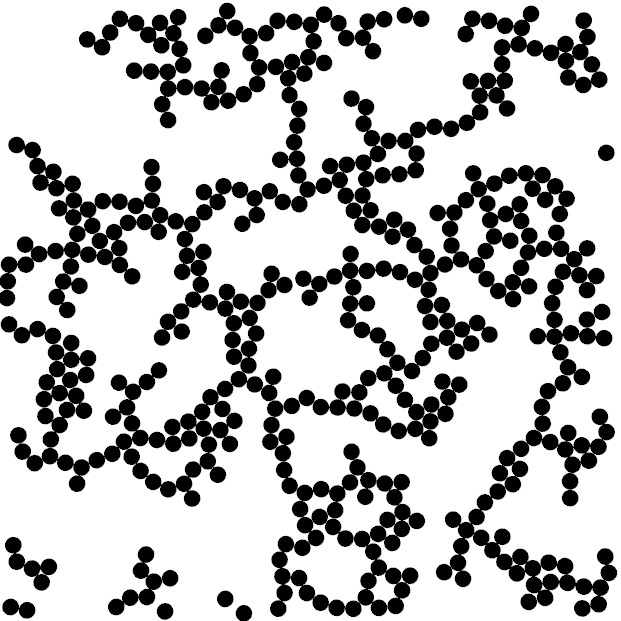}
        \caption{Minimized state ($gtol=0.01$)}
        \label{subfig:min_final}
    \end{subfigure}
    \caption{Structure of a network of aggregates (25 aggregates of 15 particles + 25 aggregates of 5 particles) prior to, and post energy minimization under van der Waals interactions}
    \label{fig:minimization Progression} 
\end{figure}

\begin{figure}[H]
    \centering
    \begin{subfigure}[b]{0.45\textwidth}
        \centering
        \includegraphics[width=\textwidth]{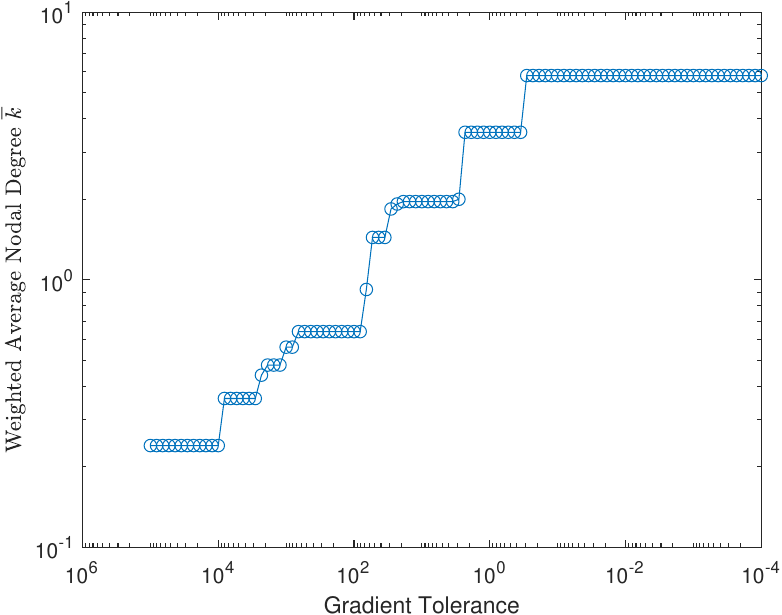}
        \caption{Gradient vs average nodal degree}
    \end{subfigure}
    \begin{subfigure}[b]{0.45\textwidth}
        \centering
        \includegraphics[width=\textwidth]{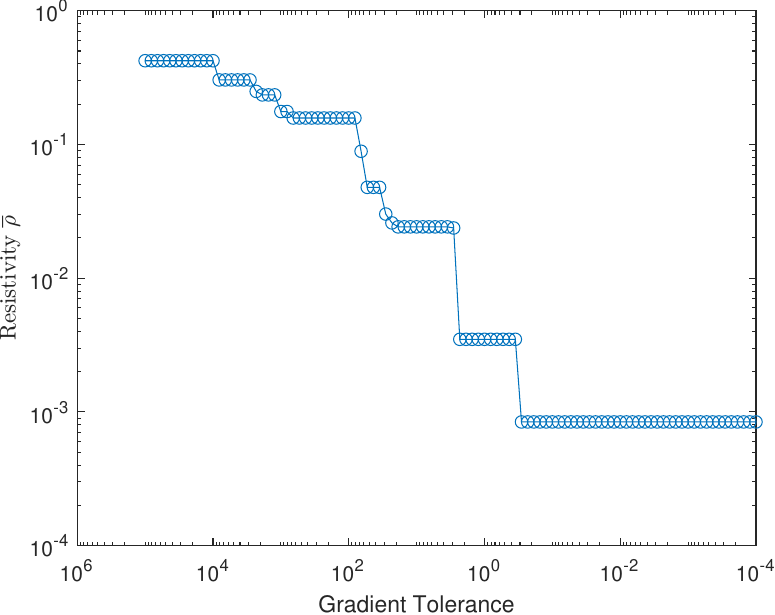}
        \caption{Gradient vs resistivity}
    \end{subfigure}
    \caption{Graph measures with minimization progress, gradient tolerance represents how close the simulation is to a local minimum (equilibrium). The effect of the van der Waals interactions is to increase network connectivity and decrease resistivity by several orders of magnitude}
    \label{fig:gradient_tol_measures}
\end{figure}

Figure~\ref{fig:minimization Progression} displays the network structure of System 1 prior to minimization, and post-minimization. Figure~\ref{fig:gradient_tol_measures} displays the conductivity and connectivity measures as the minimization proceeds, as measured by the gradient tolerance. The smaller the gradient tolerance the closer the simulation is to a local minimum (equilibrium state). These results demonstrate that the van der Waals interaction increases the connectivity and decreases the resistivity of a filler network.

\end{multicols}

\begin{multicols}{2}

Figure~\ref{fig:Sim_results} displays the results of strain simulations with three different filler densities (filler density is varied by modifying the size of the simulation region). For comparison, Figure~\ref{fig:experiments} shows experimental results using samples of Ecoflex 0045 silicone and Vulcan XC72R Carbon black, produced and tested as described in \cite{ritchie2024electromechanical}. Several key qualitative features of the piezoresistive behaviour observed in experiments are reproduced in these simulations, including long timescale resistivity relaxation, large resistivity changes under strain with fixed filler density, non-monotonic secondary resistivity peaks upon strain relaxation, and an increasing significance of the secondary peak with increasing filler density.

\end{multicols}

\begin{figure}[H]
    \centering
    \begin{subfigure}[b]{0.45\textwidth}
        \centering
        \includegraphics[width=\textwidth]{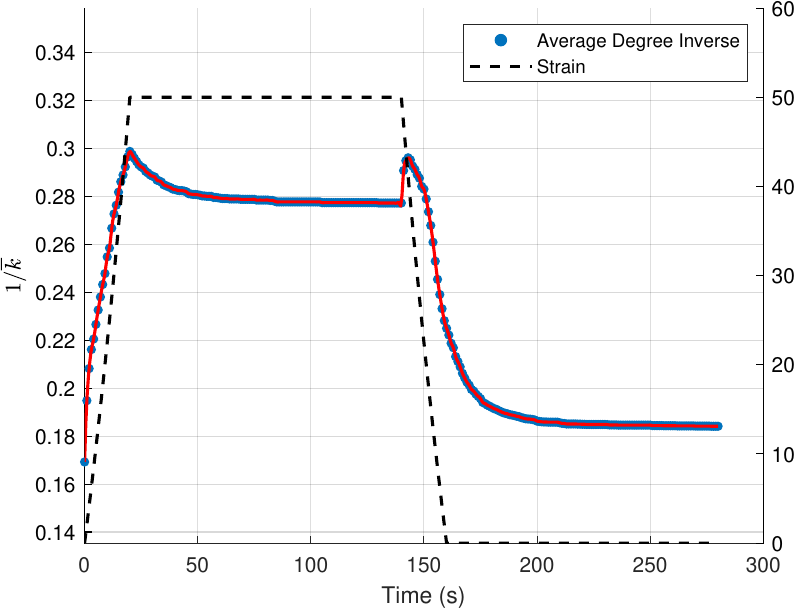}
        \caption{22\% Density - Inverse Degree}
    \end{subfigure}
    \begin{subfigure}[b]{0.45\textwidth}
        \centering
        \includegraphics[width=\textwidth]{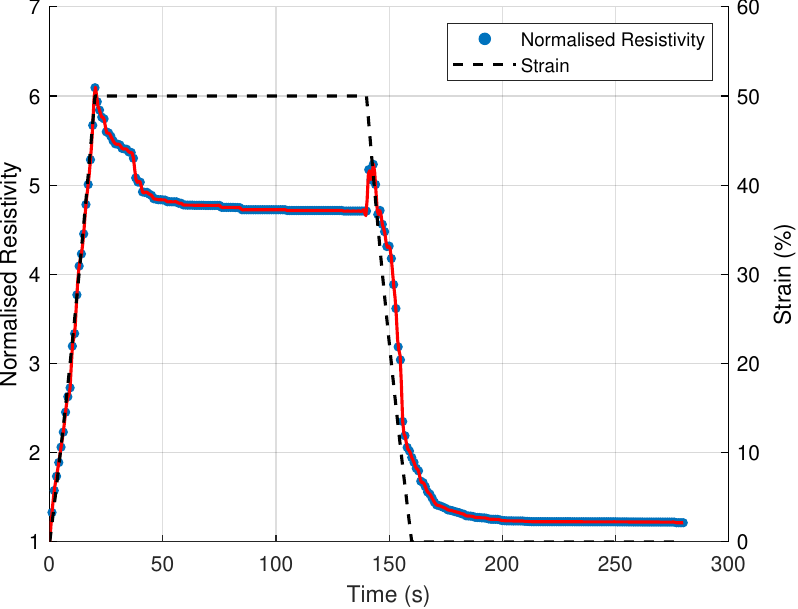}
        \caption{22\% Density - Resistivity}
    \end{subfigure}
\end{figure}
\begin{figure}[H]
    \ContinuedFloat
    \centering
    \begin{subfigure}[b]{0.45\textwidth}
        \centering
        \includegraphics[width=\textwidth]{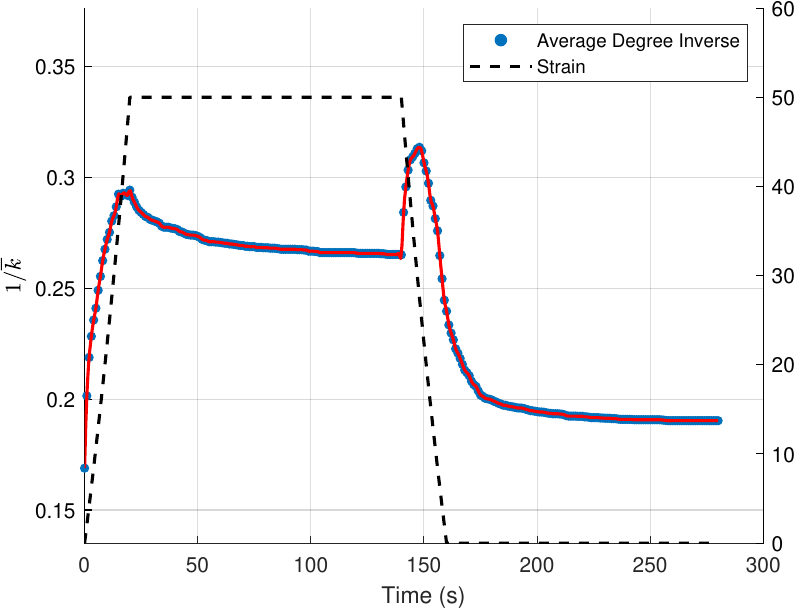}
        \caption{30\% Density - Inverse Degree}
    \end{subfigure}
    \begin{subfigure}[b]{0.45\textwidth}
        \centering
        \includegraphics[width=\textwidth]{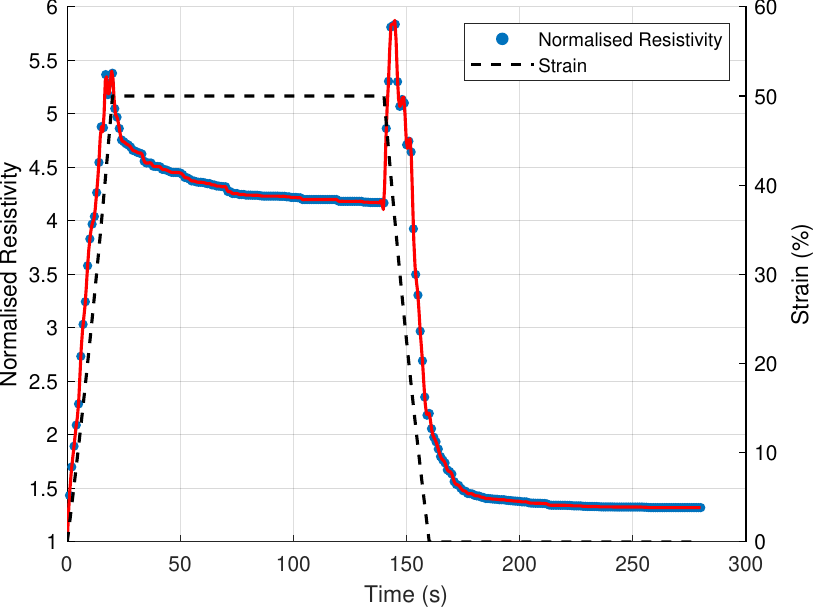}
        \caption{30\% Density - Resistivity}
    \end{subfigure}
\end{figure}
\begin{figure}[H]
    \ContinuedFloat
    \centering
    \begin{subfigure}[b]{0.45\textwidth}
        \centering
        \includegraphics[width=\textwidth]{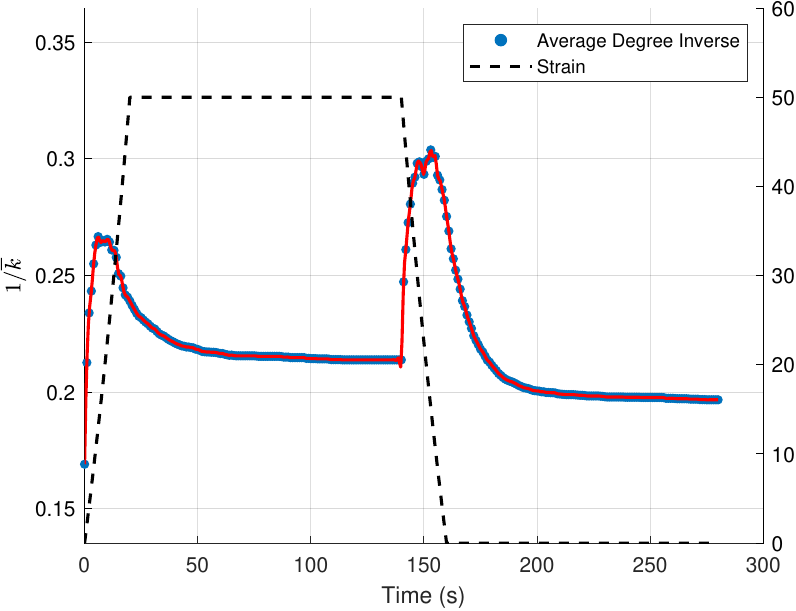}
        \caption{38\% Density - Inverse Degree}
    \end{subfigure}
    \begin{subfigure}[b]{0.45\textwidth}
        \centering
        \includegraphics[width=\textwidth]{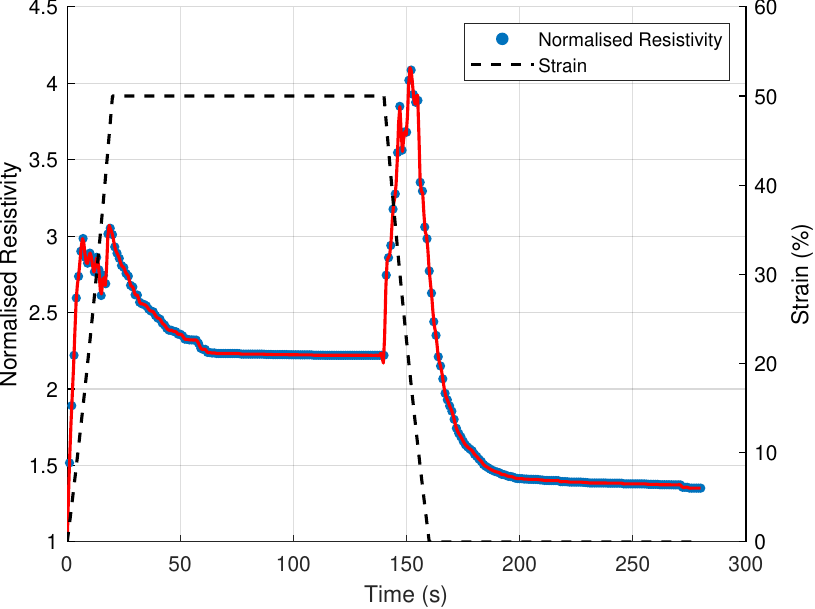}
        \caption{38\% Density - Resistivity}
    \end{subfigure}
    \caption{Simulation Results - $\frac{1}{\overline{k}}$ and $\overline{\sigma}$ for 25 aggregates of 15 particles and 25 aggregates of 5 particles at various densities under 50\% uniaxial strain. Time dependence, non-monotonic secondary peak, and increasing significance of non-monotonicity with filler density is reproduced}
    \label{fig:Sim_results}
\end{figure}
\begin{figure}[H]
    \centering
    \begin{subfigure}[b]{0.45\textwidth}
        \centering
        \includegraphics[width=\textwidth]{figures/Results/straining/experiments/ecoflex_5_percent.pdf}
        \caption{Ecoflex 0045 with 5 wt\% Vulcan XC72 Carbon Black}
    \end{subfigure}
    \begin{subfigure}[b]{0.45\textwidth}
        \centering
        \includegraphics[width=\textwidth]{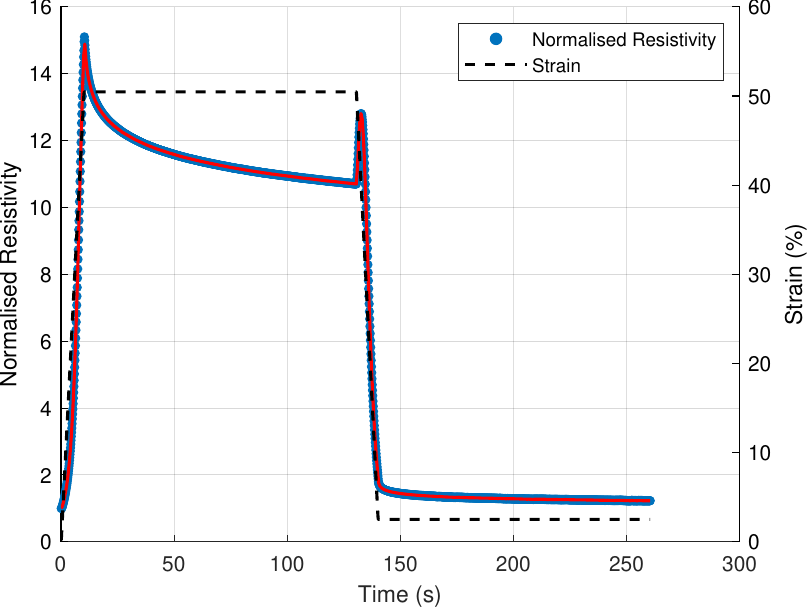}
        \caption{Ecoflex 0045 with 6 wt\% Vulcan XC72 Carbon Black}
    \end{subfigure}
\end{figure}
\begin{figure}[H]
    \centering
    \ContinuedFloat
    \begin{subfigure}[b]{0.45\textwidth}
        \centering
        \includegraphics[width=\textwidth]{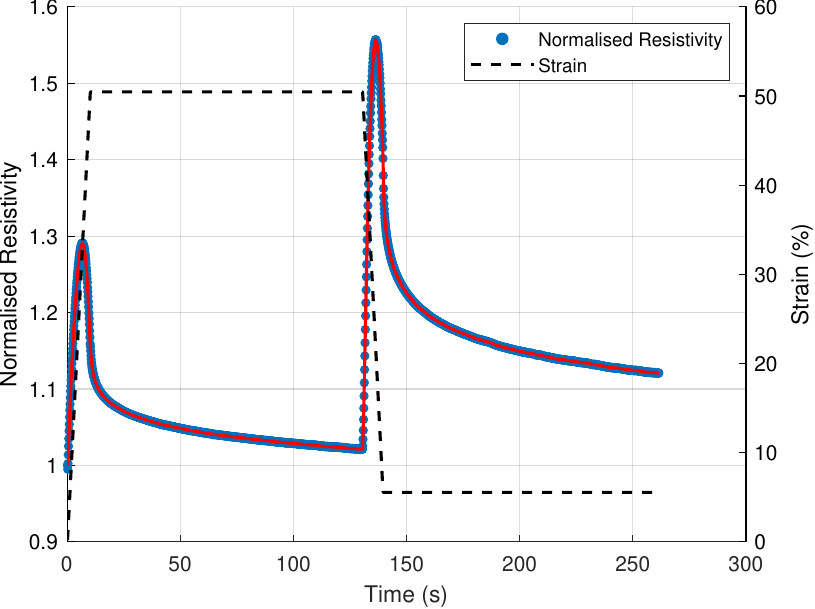}
        \caption{Ecoflex 0045 with 9 wt\% Vulcan XC72 Carbon Black}
    \end{subfigure}
    \caption{Experimental results for comparison - measured resistivity of Ecoflex 0045 with varying densities Vulcan XC72R carbon black under 50\% uniaxial strain, produced and tested as described in \cite{ritchie2024electromechanical}}
    \label{fig:experiments}
\end{figure}
\begin{multicols}{2}

\begin{figure}[H]
    \centering
    \begin{subfigure}[b]{0.45\textwidth}
        \centering
        \includegraphics[width=\textwidth]{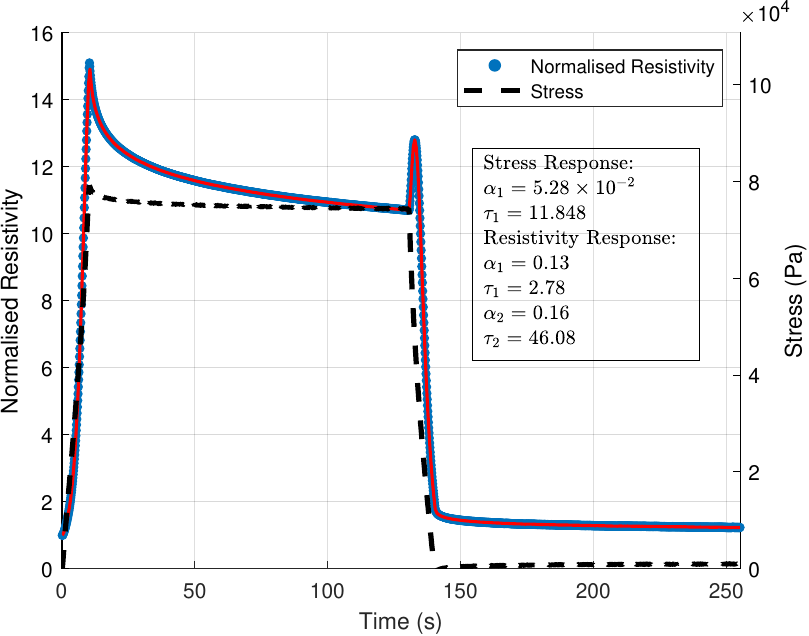}
        \caption{6 wt\% Experiment}
    \end{subfigure}
    \begin{subfigure}[b]{0.45\textwidth}
        \centering
        \includegraphics[width=\textwidth]{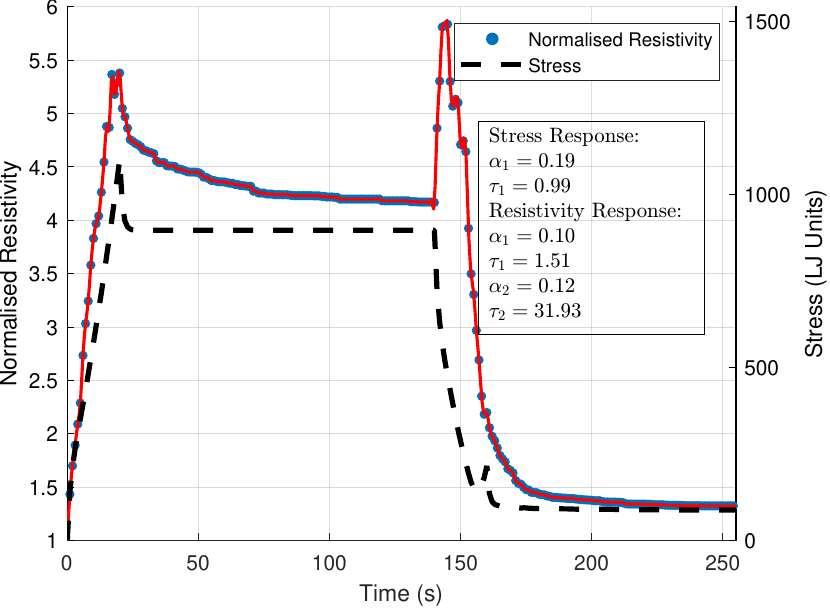}
        \caption{30\% Density Simulation}
    \end{subfigure}
    \caption{Stress and resistivity under 50\% uniaxial strain for a) experimental - Ecoflex 0045 + 6 wt\% Vulcan XC72R carbon black and b) simulation - 30\% density by area, 25 aggregates of 15 particles and 25 aggregates of 5 particles. Simulation results reproduce the different time dependent behaviour of viscoelastic stress and piezoresistivity}
    \label{fig:stress_resistivity}
\end{figure}

Figure~\ref{fig:stress_resistivity} compares the resistivity and mechanical stress profiles from the 30\% density simulation and an experimental comparison. As in the experiment, the simulation results show that the time dependent features of the piezoresistive response are distinct from the stress response, being of a larger magnitude, and operating over different timescales.
\end{multicols}
\begin{multicols}{2}

Finally, Figure~\ref{fig:dif_cond_defs} displays the results of the 30\% density simulation with different tunnelling distance thresholds, and using the Simmons tunnelling approximation. The qualitative features of the piezoresistive response are not strongly affected by the choice of tunnelling model, until the tunnelling threshold grows extremely large.
\end{multicols}

\begin{figure}[H]
    \centering
    \begin{subfigure}[b]{0.45\textwidth}
        \centering
        \includegraphics[width=\textwidth]{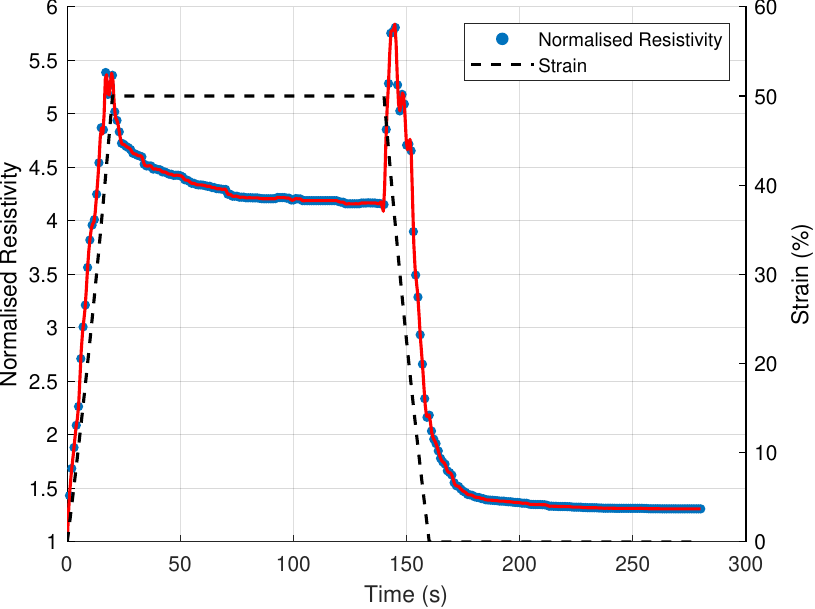}
        \caption{Threshold: 0.6 nm}
    \end{subfigure}
    \begin{subfigure}[b]{0.45\textwidth}
        \centering
        \includegraphics[width=\textwidth]{figures/Results/straining/sims/like_8_res.pdf}
        \caption{Threshold: 3 nm}
    \end{subfigure}
    
\end{figure}
\begin{figure}[H]
    \ContinuedFloat
    \centering
    \begin{subfigure}[b]{0.45\textwidth}
        \centering
        \includegraphics[width=\textwidth]{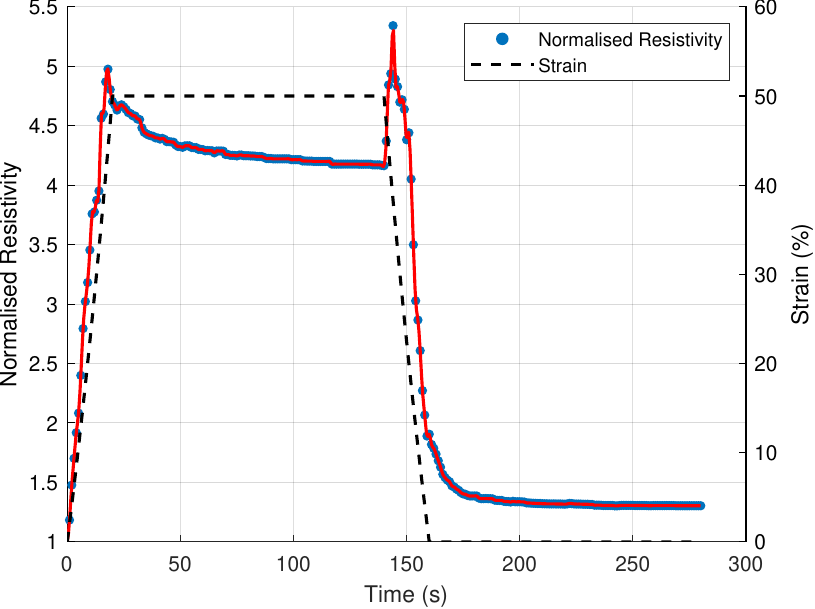}
        \caption{Threshold: 10 nm}
    \end{subfigure}
    \begin{subfigure}[b]{0.45\textwidth}
        \centering
        \includegraphics[width=\textwidth]{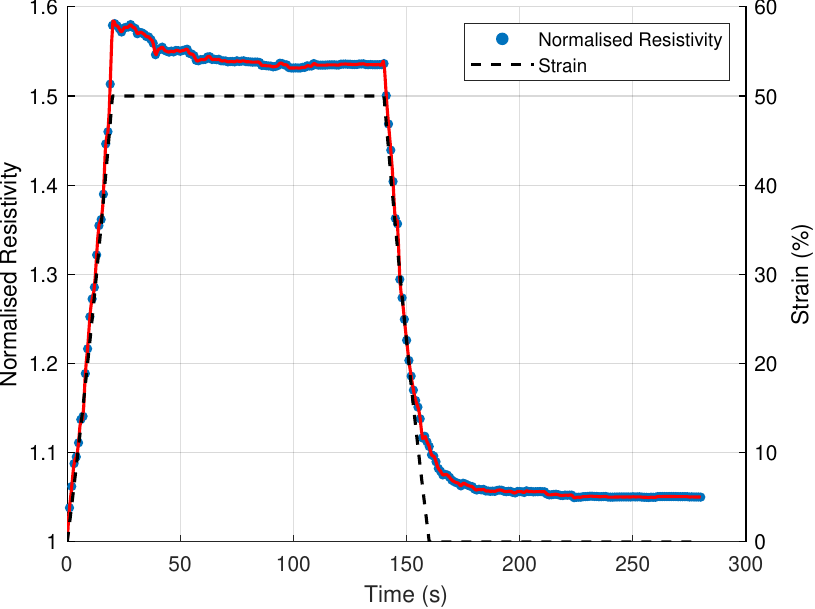}
        \caption{Threshold: 100 nm}
    \end{subfigure}
\end{figure}
\begin{figure}[H]
    \ContinuedFloat
    \centering
    \begin{subfigure}[b]{0.45\textwidth}
        \centering
        \includegraphics[width=\textwidth]{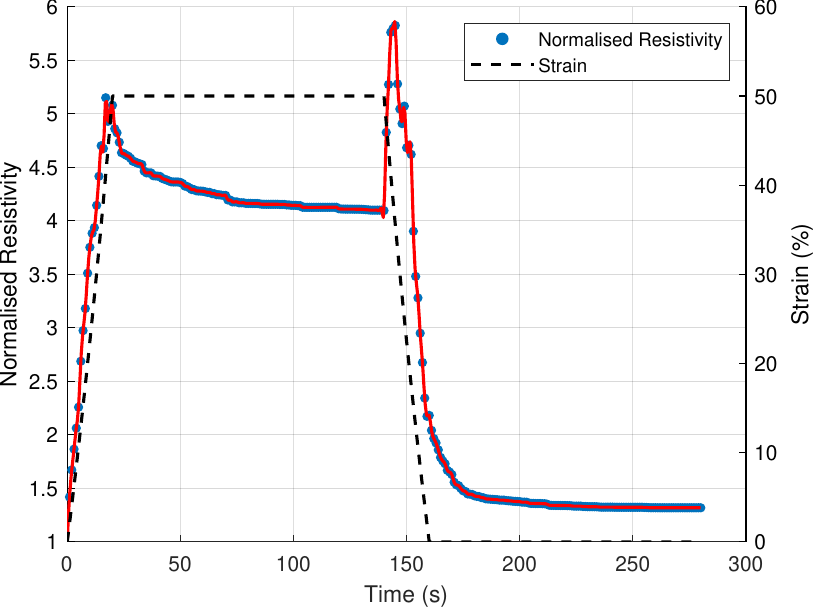}
        \caption{Simmons Approximation, $\Delta E$: 1 eV}
    \end{subfigure}
    \begin{subfigure}[b]{0.45\textwidth}
        \centering
        \includegraphics[width=\textwidth]{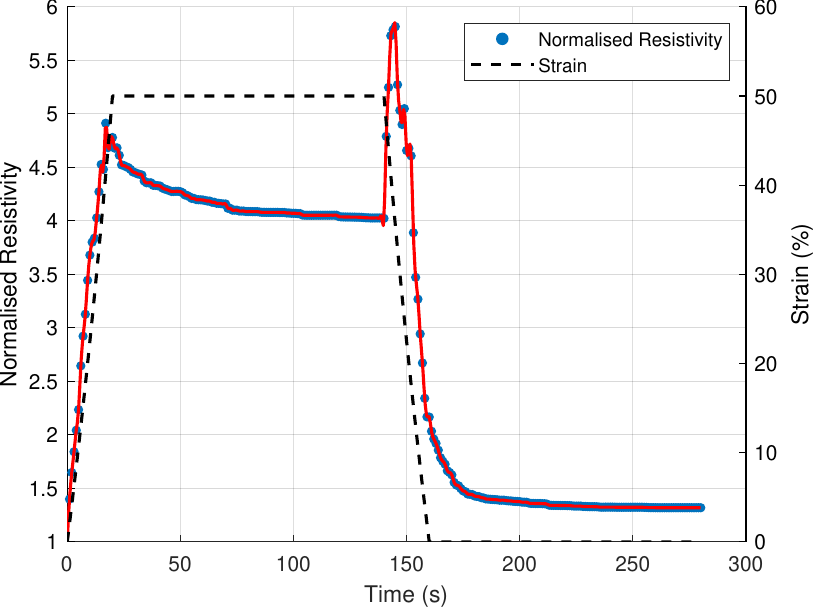}
        \caption{Simmons Approximation, $\Delta E$: 4 eV}
    \end{subfigure}
    \caption{Simulated resistivity under strain of 25 aggregates of 15 particles and 25 aggregates of 5 particles at 30\% density by area, under various different models of interparticle conductivity. Qualitative results are insensitive to exact details of conduction model}
    \label{fig:dif_cond_defs}
\end{figure}

\begin{multicols}{2}

\section{Discussion}\label{sec:discussion}
Our work supports the hypothesis that van der Waals interactions drive network formation in carbon-elastomer composites, and their competition with viscoelastic stress underlies the time-dependent piezoresistive response. 

This work demonstrated this behaviour with a custom network-based microstructural model which examined the relationship between interparticle van der Waals interactions, matrix viscoelasticity, network structure, and conductivity.

Phenomenological and microstructural models for carbon-elastomer piezoresistivity have been introduced in other works \cite{Mersch2023properties,yong2022modeling,wiessner2020piezoresistivity,Haghgoo2024breadth,WANG2018theelectro,GONG2014carbon,Haghgoo2023analytical,Haghgoo2020acomprehensive,hu2012multi,shin2023theoretical,gbaguidi2018monte}. However, in this study, several previously unexplained phenomena have been reproduced for the first time with a microstructural model, explicitly demonstrating how these anomalous behaviours can emerge from a network of interparticle interactions.
The reproduced phenomena include:
\begin{itemize}[nosep]
    \item Large resistivity changes under strain with constant filler density.
    \item Long timescale resistivity decay at constant strain.
    \item Non-monotonic secondary peak upon strain relaxation.
    \item Increasing significance of secondary peak with increasing filler density.
    \item Distinct time response of resistivity and stress.
\end{itemize}

Several assumptions and approximations were employed in this modelling. A precise value for the Hamaker constant $A$ is not known, but for qualitative inferences an approximate, typical value is sufficient. $A$ was set to $1\times10^{-19}$~J for this work, being of the order typically observed for many materials, and close to calculated values for carbon black dispersed in a variety of media \cite{HAMAKER1937thelondon,DAGASTINE2002calculations}.
As displayed in Figures~\ref{fig:minimization Progression} and \ref{fig:gradient_tol_measures}, the inclusion of van der Waals interactions has a significant impact on the network structure. In comparison to a randomly generated system of aggregates, effective resistivity drops by several orders of magnitude when compared to a random distribution. This aligns with existing results showing that increased aggregation and therefore improved conductivity is observed when carbon black is dispersed in lower viscosity solutions \cite{rwei2002dispersion}.

Assumptions were also needed when choosing the viscoelastic parameters. As the viscoelastic matrix is represented in these simulations by a two-dimensional network of one-dimensional bonds, a direct relationship between matrix properties and simulation parameters cannot yet be made. However, it can be shown with reasonable assumptions that  viscoelastic stresses in real composites are of a similar magnitude to the expected stresses generated by van der Waals forces. The derivation of this result is described in detail in the Supplementary Material. This similarity lends credence to the idea that van der Waals interactions continue to exert an influence on the material's conductivity under deformation. Thus, for this work, viscoelastic parameters were chosen for demonstrative purposes such that viscoelastic stresses in the bonds were of a similar magnitude to the maximum van der Waals attractions.

Figures~\ref{fig:Sim_results} and \ref{fig:experiments} demonstrate that the qualitative features of the simulations reproduce the complex piezoresistive profile observed in experiment, including the key result that non-monotonic behaviour increases in significance with higher filler densities. This result makes sense in light of the theory - as this secondary peak is caused by interparticle interactions, it becomes more prominent as particle density increases.

Figure~\ref{fig:stress_resistivity} demonstrates that even though the time dependence of resistivity is a result of viscoelasticity, it is a distinct phenomenon, displaying unique features such as the secondary peak, and operating over different characteristic timescales. This surprising result highlights that the complexity of this phenomenon is an emergent feature of the network of particle interactions, and cannot easily be derived from a simple analysis of material and particle properties considered in isolation.

Finally, Figure~\ref{fig:dif_cond_defs} shows that the piezoresistive profile is fairly insensitive to the precise conduction model used, only changing significantly once the tunnelling threshold grows infeasibly large. The inclusion of a more complex tunnelling model such as the Simmons approximation also does not significantly affect the results. Once again this result demonstrates that the complex piezoresistive behaviour is driven by the changes to the structure of the filler network, as opposed to resulting from a complex conduction model between individual particles.

In aggregate, these results demonstrate that the complicated piezoresistive behaviour of carbon-elastomer composites can be understood when the network of interactions is considered, and a network-based modelling methodology is introduced for this purpose. 

There are still significant limitations to this approach, however. Currently, the two-dimensional nature of the simulations, and the use of one-dimensional bonds to represent the elastomer mean that the viscoelastic properties used in the simulation are not derivable from real material properties. Tuning of simulation parameters could potentially quantitatively reproduce a range of experimental results, but would be of little explanatory use without a more clearly defined relationship between material properties and simulation parameters. With proper optimization of the minimization procedure to fully exploit the sparsity structure of the filler network, and parallelization of the algorithm, scaling to larger, three-dimensional simulations will be possible. Furthermore, to more accurately represent deformation, a transition from a one-dimensional representation of viscoelastic bonds based on bond-based peridynamics to a full deformation-tensor representation of the elastomer as in state-based peridynamics \cite{warren2009nonordinary} will likely be necessary. Until such improvements are made to the approach, the modelling methodology is confined to qualitative, explanatory insight, rather than being capable of quantitative predictions.

Further experimental investigations that characterise composite network structure under deformation, for example with small-angle x-ray scattering (SAXS) \cite{BEUTIER2022insitu}, would also be of value in further validating the modelling results.

On the other hand, the methodology presented here may be of use, even in its current form, in the explanation of other phenomena in nanocomposite materials that are dependent on filler network structure and interparticle interactions.

\section{Conclusions}
This work introduced a custom network based modelling approach to analyse the mechanism of piezoresistivity in carbon-elastomer nanocomposites. The modelling results indicate:
\begin{itemize}
    \item Van der Waals interactions play a major role in filler networking even in cured composites.
    \item Network restructuring can explain piezoresistivity, without detailed tunnelling conduction models.
    \item Time dependence of the resistance response depends on the viscoelasticity of the polymer, but exhibits novel features that arise from the network of particle interactions.
    \item Network-based modelling can help elucidate general behavioural features in complex materials.
\end{itemize}
The results highlight the potential of network/graph-based modelling to improve our understanding of structure-property relationships in complex composite materials. This improved understanding could one day be leveraged in the design of novel materials with tailored properties.

\newpage
\end{multicols}

\bibliographystyle{unsrt}

\end{document}